\documentclass{emulateapj}
\usepackage{graphicx,times}
\shorttitle{On the role of the magnetic field on jet emission in X-ray binaries}
\shortauthors{P.Casella and A. Pe'er}
\begin{document}
\title{On the role of the magnetic field on jet emission in X-ray binaries}

\author{P. Casella\altaffilmark{1} and A. Pe'er\altaffilmark{2,3}}
\altaffiltext{1}{Astronomical Institute "A. Pannekoek", University of
  Amsterdam, Kruislaan 403, 1098 SJ Amsterdam, The Netherlands; p.casella@soton.ac.uk}
\altaffiltext{2}{Space Telescope Science Institute, 3700 San Martin Dr, Baltimore, MD, 21218 }
\altaffiltext{3}{Giacconi Fellow}
\thispagestyle{empty}
\begin{abstract}
Radio and X-ray fluxes of accreting black holes in their hard state are known to correlate over several orders of magnitude. This correlation however shows a large scatter: black hole candidates with very similar X-ray luminosity, spectral energy distribution and variability, show rather different radio luminosities. This challenges theoretical models that aim at describing both the radio and the X-ray fluxes in terms of radiative emission from a relativistic jet. More generally, it opens important questions on how similar accretion flows can produce substantially different outflows. Here we present a possible explanation for this phenomenon, based on the strong dependency of the jet spectral energy distribution on the magnetic field strength, and on the idea that the strength of the jet magnetic field varies from source to source. Because of the effect of radiative losses, sources with stronger jet magnetic field values would have lower radio emission. We discuss the implications of this scenario, the main one being that the radio flux does not necessarily provide a direct measure of the jet power. We further discuss how a variable jet magnetic field, reaching a critical value, can qualitatively explain  the observed spectral transition out of the hard state.

\end{abstract}


\section{Introduction}

The existence of jets in X-ray binaries (XRBs) is today widely accepted, both because of direct observation of spatially resolved jets in a few XRBs, and because of analogy with the well known case of Active Galactic Nuclei \citep[for a review see][]{fender06}. The main signature of the presence of a jet (when not directly resolved in imaging) is the detection of radio emission, with different slopes of the radio spectrum corresponding to different types of jet structure. In the hard states of many black hole candidates (BHCs), a relatively bright, nearly-flat (in $F_{\nu}$)  radio emission is observed, which is associated with the existence of a compact steady jet \citep{fender01}. In the soft state of BHCs, on the contrary, the jet is thought no longer to be present, because of the observed highly quenched radio emission \citep[e.g.][]{tan72,corbeletal01}. Theoretical work \citep[e.g.][]{lop99,meier01} has shown how a powerful jet is indeed expected to be formed when a thick accretion flow is present (as it is considered the case in the hard state), while only a much weaker jet can be formed when the accretion flow is a thin disk (as in the soft state).

The radio flux from the jet has been shown to correlate over several orders of magnitude with the X-ray flux
in the hard state of BHCs \citep{hannikainenetal98,corbeletal03,galloetal03,galloetal06}. This indicates that the jet is strongly correlated with the accretion flow (disk and/or corona), and possibly that its emission (synchrotron and/or inverse Compton) can be significantly contributing at higher frequencies \citep[e.g.][]{markoffetal05}. 
As soon as the radio/X-ray flux correlation was discovered, the physical origin of the observed scatter was also discussed. \citet{HM04} showed how the small observed scatter \citep[once a mass-dependent correction factor is introduced, see][]{MHM03,FKM04,KFC06} might imply a similar bulk velocity for all the jets, unless they are all non-relativistic.

\begin{figure}[t]
  \hspace{0.5cm}   \includegraphics[width=6.9cm]{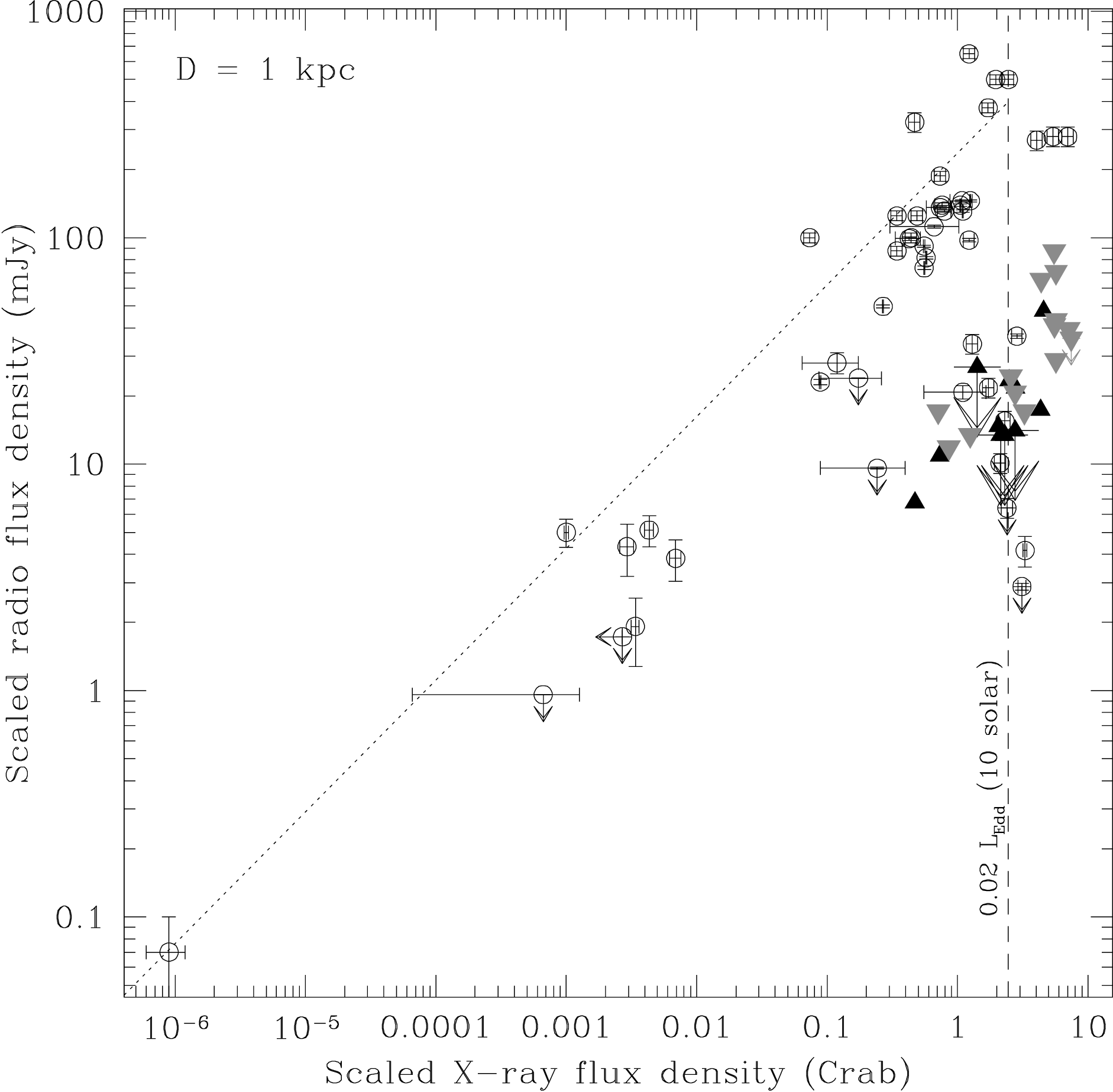}
     \caption{The radio/X-ray plane populated by BHCs in their hard state \citep[adapted from][]{gallo07}. The dotted line indicates the best fit to the correlation, as reported by \citet{gallo07}, with all the outliers lying below it. The black triangles are the outliers already reported in \citet{gallo07} (not included in the fit), the gray ones are additional measurements for SWIFT J1753.5--0127, assuming a distance of 6 kpc (from Soleri et al., in prep). No mass-correction term was applied.}
     \label{gallorel}
   \end{figure}

However in the following years, more sources have been found to lie outside the scatter of the original correlation, thus either increasing its scatter, or challenging the universality of the correlation itself. All these outliers (see Fig.~\ref{gallorel}) show a radio flux {\it below} the correlation (the so-called "radio-quiet" BHCs), and are all at high X-ray luminosities (while still in their hard state). For most of these outliers however, there are no radio measurements available at low X-ray luminosities, thus we do not know whether they remain under luminous in radio or not at lower accretion rates.

Among these radio-quiet BHCs are: XTE J1650--500 \citep{corbeletal04}, XTE J1720--318 \citep{brocksopp05,chaty06}, IGR J17497--2821 \citep{rodriguez07}, and SWIFT J1753.5--0127 \citep{cadollebel07}, although for most of them no mass estimate is available. In two of them the radio and X-ray fluxes in the hard state appear to be correlated, with a slope similar to that of the majority of BHCs but with a lower normalization:  XTE J1650--500 \citep{corbeletal04} and SWIFT J1753.5--0127 (Soleri, priv. comm.). This suggests that the main properties of their accretion flows, as for example the radiative efficiency, are similar to the main population of BHCs. This is also confirmed from the fact that both XTE J1650--500 and SWIFT J1753.5--0127, as well as the other radio-quiet BHCs, share very similar X-ray spectral properties with the rest of the population. 

The global similarities among the X-ray spectral and variability properties of all BHCs, when compared to the large scatter in radio luminosities, represents a strong challenge for broad-band jet spectral modeling. A possible obvious interpretation is that the jet emission does not contribute much to the X-ray flux. If this is the case, the radio-quiet BHCs could simply have a much weaker/fainter jet, with no effect on the X-ray emission. This would however raise fundamental questions on how similar inflows (as suggested from the X-ray properties) can lead to very different outflows (as inferred from the radio properties).
A possible alternative is that all BHCs have similar broad-band emitting jets, but with a doppler (de-)boosting factor that varies from source to source, because of different inclination angle, collimation angle and/or bulk velocity. For this interpretation to hold, one needs the X-ray and the radio fluxes to have different doppler boosting factors (i.e., a different jet speed, or a different jet angle to the line of sight, would have to affect mostly the observed radio flux, leaving mostly unchanged the X-ray flux). Also, one needs most BHCs to have mildly relativistic jets. If most jets were non-relativistic, there would not be any doppler boosting, while, if they all were highly relativistic, they would all be highly de-boosted.
Finally it is worth mentioning the possibility that some (but hardly all) radio-quiet sources actually host an accreting neutron star, which are known to be less luminous in radio \citep[at a given X-ray luminosity, albeit with a steeper correlation;][]{migliarifender06}.

In this Letter we explore a different explanation. We consider the
role of the jet magnetic field in the spectral energy distribution
from jets in XRBs, and we suggest the possibility that a
relatively small change in the magnetic field strength in the jet could
in principle account for the observed scatter in radio luminosities.  

\section{Different magnetic field regimes} \label{regimes}

In \citet{peerandcasella09} (hereinafter PC09) we presented a new model for emission from
jets in XRBs, in which electrons are accelerated only once at the base of the jet.
While we adopt the common assumption that the acceleration
process produces a power-law distribution of the energetic electrons,
based on recent theoretical models \citep[e.g.][]{spitkovsky08} we
considered both high and low energy cutoffs to the accelerated
electrons distribution. At low energies, the electrons assume a
Maxwellian distribution, while some uncertain fraction of the
electron population is accelerated to a power-law distribution at
higher energies.  Once accelerated, we considered cooling of the
electrons as they propagate along the jet (see also Kaiser 2006), via
both synchrotron radiation and possible adiabatic energy losses. The
magnetic field is assumed to decay along the jet as $1/r$, in accordance to Pointing flux conservation law.

   \begin{figure}[t]
     \includegraphics[width=8.0cm]{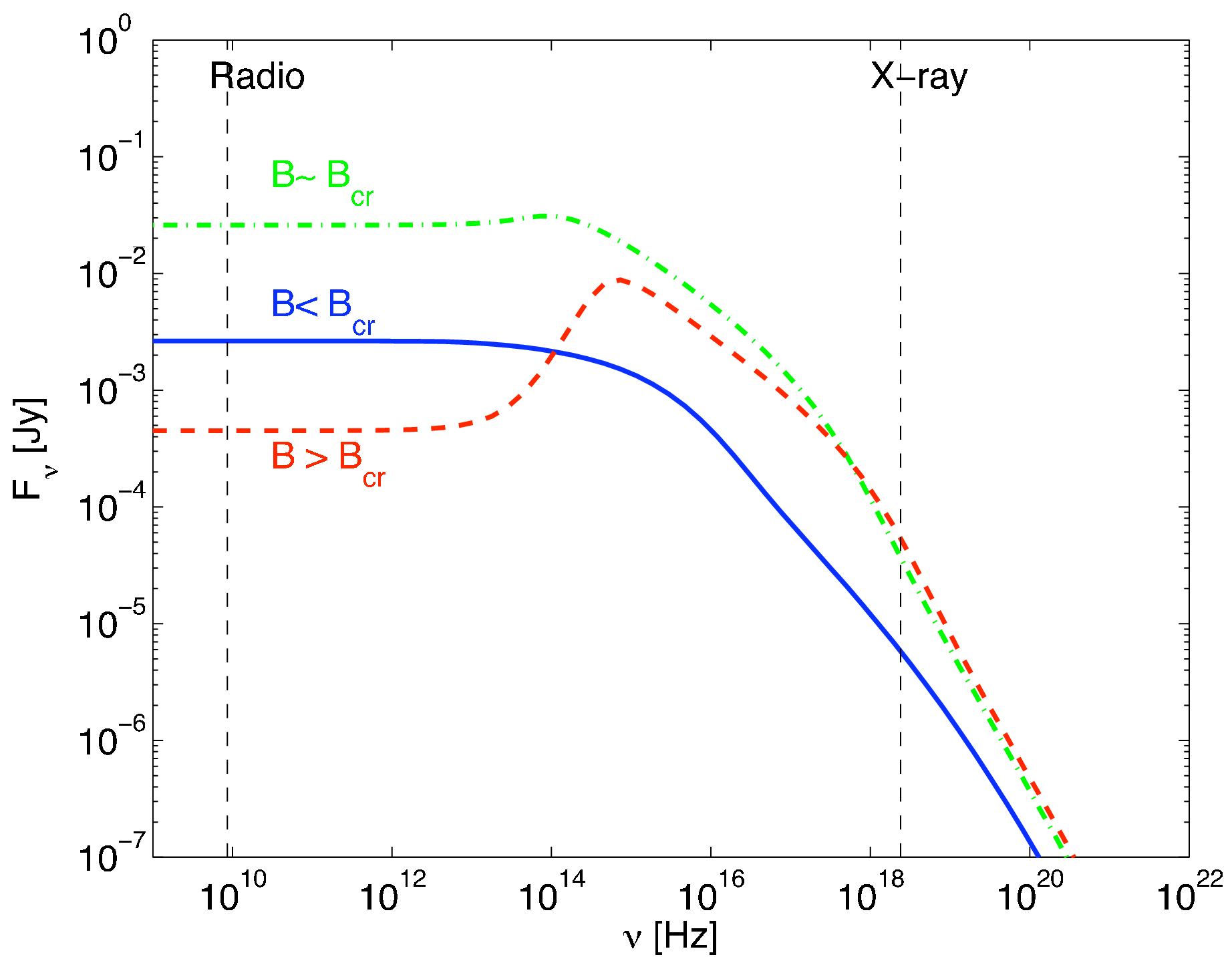}
     \caption{Spectral energy distribution from a jet, calculated for three different values of the toroidal magnetic field, for the ballistic case (no adiabatic losses considered) and an electron energy distribution with a high energy power-law tail (with spectral index $p=2.5$). 
     Two main different regimes are evident: for magnetic fields weaker than B$_{\rm cr}\sim10^5$ G (or a fraction $\sim10^{-3}$ of equipartion, for details see the text and PC09), both radio and X-ray fluxes increase with $B$, while for magnetic fields stronger than B$_{\rm cr}$ the radio flux drops and the X-ray flux saturates. The two vertical lines are drawn at 8.6 GHz and 10 keV, respectively.}
     \label{seds}
   \end{figure}


The inclusion of the low-energy cutoff in the electron distribution
implies that the observed spectrum from a jet segment has in
principle two characteristic breaks: $\nu_{peak}$, the
characteristic synchrotron emission frequency from electrons at the
peak of the Maxwellian distribution (which also corresponds to the low
end of the power-law distribution) and $\nu_{thick}$, the frequency
below which the emission is self absorbed (i.e., is in the optically
thick regime). Both frequencies depend on the characteristic energy
of the emitting electrons, as well as the magnetic
field strength. 

As we showed in PC09, for plausible parameters
that can characterise emission from XRBs,
$\nu_{peak}$  at the base of the jet is typically in the X-ray band, while
$\nu_{thick}$ is in the infrared regime. However, as both frequencies
depend on the electron energy distribution and on the strength of the magnetic
field, both cooling of the electrons and the decay of the magnetic field
along the jet act to reduce these values along the jet.

One of the most important results found in this work, is the
non-trivial dependence of the observed synchrotron spectrum on the
strength of the magnetic field. For (relatively) weak magnetic fields at the base of the jet, radiative cooling of the
electrons is insignificant. As a result, along the jet $\nu_{peak} >
\nu_{thick}$, i.e., most of the emission is always in the optically
thin regime. The observed flux is therefore dominated by emission at
$\nu_{peak}$, which, in turn, decays along the jet due to the decay of
the magnetic field. The overall observed flux (integrated along the
entire jet) is therefore relatively high below the X-ray frequency,
and in particular at radio frequencies. 
Examples of the spectra are shown in Fig.~\ref{seds} (note
though that the exact spectral shape depends on additional parameters,
such as the jet geometry, etc.; see PC09 for
detailed discussion and explanations).
  
Above a certain value of the magnetic field, which we denote as B$_{\rm cr}$, we found a qualitatively
different behaviour. If the magnetic field at the jet base exceeds a
fraction of $\sim10^{-3}$ of equipartition\footnote{For the parameters
discussed in PC09, this corresponds to a magnetic
field strength at the base of the jet of B$_{\rm cr} \sim 10^5$ G. However, 
we stress that  
the values mentioned here depend on additional parameters of the jet 
outflow, and may thus be different.\label{footo}}, then electrons rapidly cool (by synchrotron
emission) very close to the jet base. During the initial rapid
cooling, most of the emission is in the X-ray band. Therefore, the
flux at this band saturates to a constant value. The rapid cooling
implies that already close to the jet base, once the electrons cool 
below a critical energy, synchrotron emission becomes obscured, i.e.,
$\nu_{peak} < \nu_{thick}$.  If no additional heating source exists,
this situation continues along the jet: i.e., emission from electrons
that propagate along the jet remains self absorbed. At large radii the
emission is mainly at the radio band, and thus the integrated spectrum
shows a suppression of the flux at the radio band, which {\it is not}
accompanied by a similar suppression of the flux at the X-ray band. In
this regime of high magnetic field, a further increase in the strength of
the magnetic field at the jet base does not significantly change the
flux at the X-ray band, as it already saturates (the electrons radiate
most of their available energy at this band). However, emission at
radio frequencies is further suppressed, because the electrons cool to
lower energies, resulting in a further decrease in $\nu_{peak}$. 
This result may thus provide a natural explanation to the outliers seen in
Fig.~\ref{gallorel}, which show suppression of the radio flux.

For power-law distributed electrons, an additional consequence of a strong magnetic field is that the slope of the spectrum in the X-rays increases by 0.5, with respect to the spectrum obtained for
weak magnetic field (see PC09 for details).

In Fig.~\ref{seds} we plot the spectral energy distribution for three
different values of the magnetic field (B$_{\rm cr}$/30, B$_{\rm cr}$, 3B$_{\rm cr}$). The plot refers to the
ballistic case (i.e. neglecting adiabatic losses), and an electron
energy distribution with a high energy power-law tail (with spectral
index $p=2.5$), but the result of the radio quenching is general. 

\section{Discussion}\label{discussion}

As summarized in \S\ref{regimes}, a source with a jet magnetic field higher than a critical value would have a strongly quenched radio emission, but a substantially unchanged X-ray emission, with respect to sources with lower magnetic field. However, the slope of the radio/X-ray correlation is mostly set by the radiative efficiency of the accretion flow, and its dependency on the accretion rate \citep[see e.g.][for a discussion on this issue]{fenderetal03}. Therefore, 
with all the properties of the accretion flow (including its efficiency) remaining a priori unchanged, it is natural to expect for sources with strong magnetic field, a correlation between the radio and the X-ray fluxes with a similar slope to that shown by the radio-loud BHCs, but with a lower normalization.

If indeed the reason for the low radio emission in some BHCs is the stronger jet magnetic field, one can look for other possible signatures of this. An obvious direction to look at is the overall spectral shape of the jet emission. The rapid cooling of the radiating electrons implies that for magnetic field stronger than a critical value, the high-energy (thin-synchrotron) part of the spectrum steepens by a factor of 0.5 with respect to the spectrum obtained for weaker magnetic field (see Fig.~\ref{seds}).

One would thus expect the radio-quiet BHCs to show a softer X-ray spectrum, assuming the X-ray emission arises from the jet. However, such a steepening is expected already below the critical value B$_{\rm cr}$ described in \S\ref{regimes}. This means that, a priori, both radio-quiet and radio-loud BHCs might have a high-enough magnetic field to cause the X-ray steepening, thus showing a similar X-ray spectral slopes.
Moreover, and more importantly, the X-ray spectrum of BHC in their hard state appears to be different from a simple power law, as would be expected in the case of pure synchrotron emission. Indeed, the model presented in PC09 is clearly a simplification of a much more complex reality. Important contributions from processes other than synchrotron (mainly Comptonization, from a hot corona or from the jet itself) are expected to substantially modify the spectral energy distribution at high energies \citep[see e.g.][]{tom05,markoffetal05}. 
Precise estimates and predictions of the slope at high energies are thus expected to be more complex than presented in Fig.~\ref{seds}, but the main result on the role of the magnetic field holds, and is independent on the origin of the X-ray emission.

Furthermore, our model considers only a single acceleration episode. It is natural to expect the production of internal shock waves \citep{kss00}, which may further heat the particles inside the jet \cite[see e.g.][]{jamiletal08}. This would inevitably modify the results presented here, which however would qualitatively hold (albeit with higher values of B$_{\rm cr}$).

While a clear feature of the model discussed here is the peaked synchrotron emission at optical wavelengths for strong B, we expect the overlapping emission from the companion star and the accretion to make difficult to spectrally identify this feature. On the other hand, it may be possible to detect it through polarimetry studies.

\subsection{Magnetic field evolution during outbursts} \label{softstate}

Within the framework of our model, with the intensity of the magnetic field in the jet influencing the overall jet spectral emission, it is reasonable to ask how much of the observed spectral evolution during a BHC outburst could be explained in terms of changes in the magnetic field in the jet of a single source. In this perspective, the usual conclusion that the jet is switched off when the radio is quenched does not necessarily hold any longer. More generally, according to our model, the radio (or infrared) flux cannot any longer be considered a good tracer of the jet power.

For example, radio emission from BHCs is known to quench at (or around) the transition out of the hard state \citep[e.g. the well studied GX 339--4; for a discussion on more sources see][]{fenderetal09}. Could this radio quenching be due to an increase in the magnetic field of the jet? If the magnetic field were to increase during the hard-state rise of the outburst, the effects on the radio/X-ray correlation would not be distinguishable from those due to the increase of the accretion rate. In both cases the overall jet spectrum would remain approximately the same, albeit increasing its normalization (under the hypothesis of the X-rays coming from the jet). However, once the magnetic field reaches a critical value, a qualitative change would occur. A further increase of the magnetic field would leave almost unchanged the X-ray luminosity, which for a given accretion rate would saturate, although the X-ray spectrum would show a steepening. At the same time the radio luminosity would drop. This is qualitatively consistent with what is observed during the transition out of the hard state: the X-ray luminosity remains approximately constant, the X-ray hardness shows a sharp turnoff, and the radio-infrared emission quenches \citep[e.g.][]{homanetal05}. The X-ray luminosity at which this spectral transition happens, would then depend on the accretion rate, allowing one source to go through the spectral transition at different luminosities in different outbursts (as in the case of GX 339--4). However, GX 339--4 itself has been found to show parallel tracks in the radio/X-ray plane, with the hard-state luminosities from two different outbursts being correlated with the same slope but a factor of $\sim$2 difference in normalization \citep[][Corbel et al., in prep.]{nowaketal05}. This would require, in the context described in this Letter, for the jet magnetic field in GX 339--4 be always at or above the critical value.

Little is known about radio emission in the soft state of BHCs, with only a small number of sources being detected \citep[see e.g.][and references therein]{fenderetal09}. In XTE J1650-500, the radio spectrum in the soft state is consistent with being flat or inverted, thus consistent with thick synchrotron emission from a jet. However, an interaction of previously ejected matter with the ISM seems more plausible \citep{corbeletal04,fenderetal09}.

In order to further test this scenario, intensive and sensitive broad-band monitoring of  BHC outbursts will be needed, as to track the full spectral evolution, from radio to X-rays, across all spectral transitions.

\subsection{Upon what the jet magnetic field depends on}\label{magnetic}

A discussion on the origin of a (variable) jet magnetic field in BHCs is beyond the aims of this Letter. Here we limit ourselves to a few general considerations. The intensity of the magnetic field in the jet is one of the several unknowns of jet physics. It is often assumed to be at equipartition with the electrons energy, although this is not always the case [e.g., Poynting-flux dominated jets are widely discussed in the literature \citep[e.g.][and references therein]{lovelaceetal02}]. In the model discussed here, low values of the magnetic field are needed, which are not easy to reconcile with the generally recognized need for strong magnetic fields to launch the jet itself \citep[e.g.][]{BP82}. However, models predict that the magnetic field {\it in the disk} affects the jet power \citep[e.g.][]{meier01}, but no conclusions have been drawn about the remaining magnetic field in the jet. Furthermore, the existence of additional sources of particles heating might result in increasing the values considered here.

No detailed studies have been performed on the dependencies of the magnetic field in the jet on the black-hole spin, or on the accretion flow properties. For example, very little or nothing is known about how a possible misalignment between the accretion disk rotation axis and the BH spin would affect the jet properties, and in particular the magnetic field. A variable advection of magnetic field through the disc \citep[e.g.,][]{taggeretal04,rothstein08}, for example because of different magnetic properties of the accreted matter, might in principle also result in a variable jet magnetic field.

\section{Conclusion} \label{conclusion}

We have discussed the potentially important role played by the jet magnetic field in describing the observed spectral behaviour of BHCs. In particular, we showed how a scatter of the jet magnetic field values could be the cause for the observed scatter in the radio/X-ray luminosity correlation shown by BHCs: sources with a stronger jet magnetic field, above a critical value, would have a lower radio luminosities. Furthermore, we discussed how the observed spectral transition out of the hard state can be qualitatively explained by a jet magnetic field reaching a critical value. This would cause a saturation of the X-ray luminosity, a relatively sharp turnoff of the X-ray hardness, and a quenching of the radio-to-infrared jet emission. More generally, we have discussed how  the radio can no longer be considered a good tracer of the jet power. This implies that the usual conclusion that the jet is switched off when the radio is quenched, does not necessarily hold any longer. This conclusion is general, and might hold also for other types of sources, as Active Galactic Nuclei and accreting neutron stars. In particular, the strong dipolar magnetic field of the neutron star might give an important contribution to the initial radiative losses, if the launching region is close enough to the compact object.


\acknowledgments

We would like to thank D. Maitra, D. Russell, O. Jamil, P. Soleri, C. Fragile, S. Markoff, S. Corbel, R. Fender, and T. Maccarone for very useful comments and discussions, and to E. Gallo for providing the original data for Fig.~\ref{gallorel}. PC is particularly grateful to G. Ghisellini for an inspiring conversation on jet physics and to P. Soleri for allowing us to plot his unpublished data of SWIFT J1753.5-0127. We thank the referee for detailed and constructive comments. This work was partially supported by an NWO Spinoza grant to M. van der Klis. AP is supported by the Riccardo Giacconi Fellowship award of the Space Telescope Science Institute.


\end{document}